\documentclass[12pt]{article}
\newcommand{\be}{\begin{equation}}
\newcommand{\ee}{\end{equation}}
\newcommand{\bea}{\begin{eqnarray}}
\newcommand{\eea}{\end{eqnarray}}
\newcommand{\bt}{\begin{tabular}}
\newcommand{\et}{\end{tabular}}
\newcommand{\ba}{\begin{array}}
\newcommand{\ea}{\end{array}}
\newcommand{\ov}{\overline}
\newcommand{\wt}{\widetilde}
\newcommand{\bvec}{\mathbf}

\def\ne{\hbox{$\nu_e \!$ }}
\def\nm{\hbox{$\nu_\mu \!$ }}
\def\nt{\hbox{$\nu_\tau \!$ }}
\def\nel{\hbox{$\nu_{eL} \!$ }}
\def\nml{\hbox{$\nu_{\mu L} \!$ }}
\def\ncel{\hbox{$\nu_{e L}^c \!$ }}
\def\ncml{\hbox{$\nu_{\mu L}^c \!$ }}

\def\rt{\hbox{$\rightarrow$ }}

\begin{document}
\setcounter{page}{0}
\thispagestyle{empty}
\baselineskip=20pt
%------------------------------------------------------------------

\hfill{
\begin{tabular}{l}
DSF$-$98/6 \\
INFN$-$NA$-$IV$-$98/6 \\
hep-ph/9803471
\end{tabular}}

\bigskip\bigskip

\begin{center}
\begin{huge}
{\bf Flavour transitions of Dirac-Majorana neutrinos }
\end{huge}
\end{center}

\vspace{2cm}

\begin{center}
{\Large
Salvatore Esposito 
\footnote{e-mail: sesposito@na.infn.it}
and Nicola Tancredi 
\footnote{e-mail: tancredi@na.infn.it} \\} 
\end{center}

\vspace{0.5truecm}

\normalsize
\begin{center}
{\it
\noindent
Dipartimento di Scienze Fisiche, Universit\`a di Napoli ``Federico 
II''\\
Mostra d'Oltremare Pad. 19, I-80125 Napoli Italy \\
and \\
Istituto Nazionale di Fisica Nucleare, Sezione di Napoli\\
Mostra d'Oltremare Pad. 20, I-80125 Napoli Italy }
\end{center}

\vspace{3truecm}

\begin{abstract}
\noindent
From a phenomenological point of view, we study active-active and 
active-sterile flavour-changing (and flavour-conserving) oscillations of 
Dirac-Majorana neutrinos both in vacuum and in matter. The general 
expressions for the transition probabilities in vacuum are reported. 
We then investigate some interesting consequences following from 
particular simple forms of the neutrino mass matrices, and for the 
envisaged scenarios we discuss in detail neutrino propagation in 
matter. Special emphasis is given to the problem of occurrence of resonant 
enhancement of active-active and active-sterile neutrino oscillations 
in a medium. The peculiar novel features related to the Dirac-Majorana 
nature of neutrinos are particularly pointed out.
\end{abstract}

\vspace{1truecm}
\noindent

\newpage

\section{Introduction}

Today we have several indications in favour of non zero neutrino 
masses and mixing.\\
The solar neutrino problem, i.e. the observed deficit of solar 
neutrino fluxes \cite{SNP}, is a well established tool whose 
resolution requires (almost without doubt \cite{doubt}) neutrino 
physics beyond the (minimal) Standard Model. The acceptable solutions 
to this problem, in terms of vacuum \cite{Vsol} or matter \cite{Msol} 
flavour oscillations or spin and flavour oscillations \cite{SFsol} as 
well as in terms of active-sterile neutrino conversion \cite{Stsol}, 
all need non vanishing neutrino masses and mixing 
\cite{Vac, MSW, Akh, sterile}.\\
The second indication in favour of neutrino oscillations come from the 
observed deficit of atmospheric muon neutrinos with respect to 
electron neutrinos \cite{atmo} that can be explained in terms of \nm 
\rt \nt or \nm \rt \ne or even active-sterile neutrino transitions 
\cite{atsol}.\\
Laboratory direct searches for massive neutrinos only give, at 
present,  upper limits on neutrino masses \cite{masses} and the same 
is valid for reactor and accelerator  neutrino oscillations 
experiments \cite{reactor}, except for the LSND experiment \cite{LSND} 
whose results seem to be explained in terms of $\ov{\nm} \rt \ov{\ne}$ 
oscillations.\\
Hints for massive neutrinos also come from cosmology, looking at \nt 
as the most probable candidate for the hot component of the dark 
matter \cite{HDM} and from the observed peculiar velocities of pulsars 
\cite{Segre}. On the other hand, in Grand Unified Theories, which 
attempt to give a unified view of electroweak and strong interactions, 
massive neutrinos are predicted \cite{Buccella} together with other 
phenomena violating both lepton numbers and baryon number (such as, 
for example, proton decay).

However, the most intriguing fact is that a simple scenario with only 
three massive neutrinos cannot account for the solar neutrino problem, 
the atmospheric neutrino anomaly and the LSND result. This is because 
the three squared masses differences $\Delta m^2$ for the three 
oscillation solutions to these problems are all distinct between them: 
the resonant MSW solution to the solar neutrino problem requires 
$\Delta m^2 \, \sim \, 10^{-5}$ eV$^2$, while for the atmospheric 
anomaly $\Delta m^2 \, \sim \, 10^{-2}$ eV$^2$ is needed and 
$\Delta m^2 \, \sim \, 1$ eV$^2$ for the LSND result. Many analyses 
\cite{analyses} have been conducted for giving a unified view 
of the three problems in terms of neutrino oscillation (taking into 
account also the limits from laboratory experiments) and a coherent 
picture seems to emerge with four massive neutrinos, namely the three 
known neutrinos plus a sterile neutrino. Note that four neutrino mass 
eigenstates are needed, but not necessarily more than three neutrino 
flavour eigenstates. This scenario is easily realized if one considers 
neutrinos as Dirac-Majorana particles described by the following 
general mass term in the electroweak lagrangian \cite{sterile}:
\be
-{\cal L}^{DM}_{m} \; = \;
\sum_{l,l^{\prime}} 
\ov{\nu}_{l^{\prime}R} 
\; M^{D}_{l^{\prime}l} \;
\nu_{l^{\prime}L} \; + \;
\frac{1}{2} \;
\sum_{l,l^{\prime}} 
\ov{\nu}^{c}_{l^{\prime}R} \; 
M^{1}_{l^{\prime}l} \;
\nu_{l L} \; + \;
\frac{1}{2} \;
\sum_{l,l^{\prime}} 
\ov{\nu}^{c}_{l^{\prime}L} 
\; M^{2}_{l^{\prime}l} \;
\nu_{l R} \; + \; h.c.
\label{11}
\ee
Here $l,l^{\prime} = e, \mu , \tau$ label the three flavour 
eigenstates and $M_D$, $M_1$, $M_2$ are the Dirac and the two Majorana 
mass matrices which, in general, are hermitian and non diagonal 
(however, $M_1$ and $M_2$ have to be symmetric). To construct the mass 
term in (\ref{11}) we need the three known left-handed neutrinos (and 
their antiparticles) and other three right-handed sterile neutrinos 
(and their antiparticles) \footnote{Obviously, the generalization to 
more than three families is possible and straightforward}. After the 
diagonalization of (\ref{11}) we can obtain in general six mass 
eigenstates which are Majorana fields; so in this framework we can 
easily endow the above scenario with four massive neutrinos coming 
from the experiments.\\
Note that if neutrinos are really described by the mass term in 
(\ref{11}), the total lepton number is no longer conserved and 
peculiar phenomena, as neutrinoless double beta decay and 
neutrino-antineutrino oscillations can take place.\\
We stress that (\ref{11}) is predicted in many GUTs \cite{Buccella} in 
which the popular ``seesaw'' mechanism \cite{seesaw} can give rise to 
very small neutrino masses in a very natural way by supposing $M_1 
\approx 0$ and $M_D \ll M_2$. However, this is not the only 
possibility; recently some models assuming  $M_1 \simeq M_2$ have been 
proposed \cite{equal} for accounting the three experimental 
indications on neutrino oscillations discussed above. Here we further 
explore this last scenario and study flavour transitions of 
Dirac-Majorana neutrinos from a completely phenomenological point of 
view, adopting no particular model. This work is a generalization to 
flavour transitions of previous papers \cite{TE, previous} in which 
we studied peculiar oscillations of Dirac-Majorana neutrinos. We now 
assume, for simplicity, only two flavours, so $M_D$, $M_1$, $M_2$ in 
(\ref{11}) are $2 \times 2$ matrices in the flavour space. In the 
following section, the basic vacuum oscillations allowed by (\ref{11}) 
are studied and transition probabilities are explicitly given in the 
general case. Some very interesting consequences due to particularly 
simple forms of the mass matrix are also investigated. In section 3, 
given the effective hamiltonian of Dirac-Majorana neutrinos 
interacting with a medium, resonant matter oscillations are considered 
along with a qualitative discussion of the phenomenon with the aid  of 
the level crossing diagram. Finally, in section 4, there are our 
conclusions and remarks.

\section{Dinamical evolution of Dirac-Majorana neutrinos in vacuum}

Let us consider the propagation in vacuum of Dirac-Majorana neutrinos 
with 4-momentum $k^{\mu} \, = \, (\omega, \bvec{k})$ described by the 
following lagrangian:
\be
{\cal L} \; = \; \left( \ov{\nu} \; , \; \ov{\nu}^C \right) 
                 \left( \ba{cc} \not{k}  & 0 \\
                             0  & \not{k}
                 \ea  \right) \left( \ba{c}
                          \nu \\
                          \nu^C
                          \ea \right)
       - \,   \left( \ov{\nu} \; , \; \ov{\nu}^C \right) 
                 \left( \ba{cc} M_D  & M_M \\
                             M_M  & M_D
                 \ea  \right) \left( \ba{c}
                          \nu \\
                          \nu^C
                          \ea \right)
\label{21}
\ee
(for convenience we have absorbed a factor 1/2 in $M_D$ and $M_M$ with 
respect to the mass terms in (\ref{11})). In the chiral Weyl basis for 
the Dirac gamma matrices, denoting
\be
\nu \; = \; \left( \ba{c}
                   \nu_L \\
                   \nu_R  \ea \right) \;\;\;\;\;\;\;\;\;\;\;\;\;\;\;
\nu^c \; = \; \left( \ba{c}
                   \nu^c_L \\
                   \nu^c_R    \ea \right)
\label{22}
\ee
we have
\be
\left(  \ba{cccc}
k & 0 & M_D & M_M \\
0 & k & M_M & M_D \\
M_D & M_M & - \, k & 0 \\
M_M & M_D & 0 & - \, k \ea \right) \, 
\left( \ba{c}  \nu^c_L \\
               \nu_L \\
               \nu^c_R \\
               \nu_R \ea \right) \; = \; \omega \,
\left( \ba{c}  \nu^c_L \\
               \nu_L \\
               \nu^c_R \\
               \nu_R \ea \right) 
\label{23}
\ee
In a more compact form, indicating
\be
n_L \; = \; \left( \ba{c}
                   \nu_L \\
                   \nu_L^c  \ea \right) \;\;\;\;\;\;\;\;\;\;\;\;\;\;\;
n_R \; = \; \left( \ba{c}
                   \nu_R \\
                   \nu^c_R    \ea \right)
\ee
and 
\be
M \; = \; \left( \ba{cc} M_D & M_M \\ M_M & M_D \ea \right)
\label{25}
\ee
the equation (\ref{23}) can be written as
\be
\left( \ba{cc} k & M \\ M & - \, k \ea \right) \left( \ba{c} n_L \\ 
n_R \ea \right) \; = \; \omega \, \left( \ba{c} n_L \\ 
n_R \ea \right)
\label{26}
\ee
Eqs. (\ref{26}) shows that, in general, chirality-changing transitions 
are possible (in fact, in vacuum, chirality is not in general 
conserved), but these are suppressed with respect to the 
chirality-preserving ones in the ultrarelativistic limit, because in 
this limit chirality almost coincides with helicity, which is strictly 
conserved \cite{previous}. Due to this suppression, chirality-changing 
transitions are not of very practical interest, and we will not 
consider further these processes. At the leading order in the 
ultrarelativistic limit ($\omega - k \ll 2 k$, from (\ref{26}) we 
deduce that
\be
n_R \; \simeq \; \left( 1 \; - \; \frac{\omega \, - \, k}{2 \, k} 
\right) \, \frac{M}{2 \, k} \: n_L
\ee
while the approximate equation for $n_L$ is
\be
H \, n_L \; = \; \omega \, n_L
\label{28}
\ee
with
\be
H \; \simeq \; k \; + \; \frac{M^2}{2 \, k}
\label{29}
\ee
and the matrix $M$ is given in (\ref{25}). Here, for simplicity, we 
limit ourselves to only two flavours, for example $e$ and $\mu$, so 
that the mass matrices in (\ref{25}) have in general (assuming CP 
conserved) the following non diagonal form:
\be
M_D \; = \; \left( \ba{cc} m^D_{ee} & m^D_{e \mu} \\ m^D_{e \mu} & 
m^D_{\mu \mu} \ea \right) \;\;\;\;\;\;\;\;\;\;\;\;\;\;\;
M_M \; = \; \left( \ba{cc} m^M_{ee} & m^M_{e \mu} \\ m^M_{e \mu} & 
m^M_{\mu \mu} \ea \right)
\label{210}
\ee
The $4 \times 4$ hamiltonian in (\ref{29}) can be easily 
block-diagonalized by using the unitary matrix
\be
V \; = \; \frac{1}{\sqrt{2}} \; \left( \ba{cc} I & I \\ - \, I & I \ea 
\right)
\label{211}
\ee
where $I$ is the $2 \times 2$ identity matrix. In fact, introducing 
the Majorana states $\wt{n_L} \, = \, V \, n_L$,
\be
\left( \ba{c} \wt{n_+} \\ \wt{n_-} \ea \right) \; = \; V \: \left( 
\ba{c} \nu_L \\ \nu_L^c \ea \right) \; = \; \frac{1}{\sqrt{2}} \, 
\left( \ba{c} \nu_L \, + \, \nu_L^c \\ - \, \nu_L \, + \, \nu_L^c \ea 
\right)
\label{212}
\ee
we can rewrite (\ref{28}) as
\be
\wt{H} \, \wt{n_L} \; = \; \omega \, \wt{n_L}
\label{213}
\ee
with
\be
\wt{H} \; = \; k \; + \; \frac{\wt{M}^2}{2 \, k} \; = \; \left( 
\ba{cc} k \; + \; \frac{M_+^2}{2 \, k} & 0 \\
0 & k \; + \; \frac{M_-^2}{2 \, k} \ea \right)
\label{214}
\ee
where
\be
M_{\pm} \; =  \; M_D \; \pm \; M_M
\label{215}
\ee
It is now a simple task to completely diagonalize (\ref{214}) by means 
of the mixing matrices
\be
U_{\pm} \; = \; \left( \ba{cc}
                 \cos \, \theta_\pm & \sin \, \theta_\pm \\
                 - \sin \, \theta_\pm & \cos \, \theta_\pm
                 \ea \right)
\label{216}
\ee
with the mixing angles given by
\be
\tan \, 2 \theta_\pm \; = \; \frac{2 \, m^\pm_{e \mu}}{m_{e e}^\pm \, - 
\, m_{\mu \mu}^\pm}
\label{217}
\ee
($m^\pm_{e e} \, = \, m^D_{e e} \, \pm \, m^M_{e e}$ and so on). 
Indicating with
\bea
m_1^\pm & = & \frac{1}{2} \, \left( m_{e e}^\pm \, + 
\, m_{\mu \mu}^\pm \; + \; \sqrt{(m_{e e}^\pm \, - 
\, m_{\mu \mu}^\pm )^2 \, + \, 4 (m_{e \mu}^\pm)^2} \right)
\label{218}  \\
m_2^\pm & = & \frac{1}{2} \, \left( m_{e e}^\pm \, + 
\, m_{\mu \mu}^\pm \; - \; \sqrt{(m_{e e}^\pm \, - 
\, m_{\mu \mu}^\pm )^2 \, + \, 4 (m_{e \mu}^\pm)^2} \right)
\label{219}
\eea
the eigenvalues of the mass matrices in (\ref{215}), the four energy 
eigenvalues are
\be
E^\pm_{1,2} \; = \; k \; + \; \frac{(m_{1,2}^\pm)^2}{2 \, k}
\label{220a}
\ee
to which correspond the eigenstates
\be
\left( \ba{c} \nu_+ \\ \nu_- \ea \right) \; = \; U \, 
\left( \ba{c} \wt{n_+} \\ \wt{n_-}\ea \right) \; = \; \left( \ba{cc} U_+ & 0 
\\ 0 & U_- \ea \right) \, \left( \ba{c} \wt{n_+} \\ \wt{n_-} \ea \right) \; = \;
U \, V \, \left( \ba{c} \nu_L \\ \nu_L^c \ea \right) 
\label{221a}
\ee
where we have used  the shorthand notation
\be
\nu_\pm \; = \; \left( \ba{c} \nu_{1 \pm} \\ \nu_{2 \pm} \ea \right)
\label{222a}
\ee
Given the relation (\ref{221a}) between the energy eigenstates and the 
flavour ones, we easily get the time evolution of the states created 
by weak interactions:
\be
\left( \ba{c} | \, \nu_{eL} (t) > \\
| \, \nu_{\mu L} (t) > \\ | \, \nu_{eL}^c (t) > \\
| \, \nu_{\mu L}^c (t) > \ea \right) \; = \; V^T \, U^T \, {\cal A} \, 
U \, V \, \left( \ba{c} | \, \nu_{eL} (0) > \\
| \, \nu_{\mu L} (0) > \\ | \, \nu_{eL}^c (0) > \\
| \, \nu_{\mu L}^c (0) > \ea \right) 
\label{223a}
\ee
where
\be
{\cal A} \; = \; diag \left\{ e^{- 1 \, E_{1+} \, t}, e^{- 1 \, E_{2+} \, t}, 
e^{- 1 \, E_{1-} \, t}, e^{- 1 \, E_{2-} \, t} \right\}
\label{224a}
\ee
The transition probabilities $P(\nu_i \rt \nu_j)$ are then given by 
the squared modulus of the corresponding matrix elements $| \, < \nu_j 
(0) \, | \, \nu_i (0) > \, |^2$ in (\ref{223a}); after some 
calculations we obtain
\bea
P(\nel \rightarrow \nml) & = &\frac{1}{4} \left( 
 \sin^2 2 \theta_+ \, \sin^2 \frac{\Delta m^{2}_{+}}{4k} t \; + \;
 \sin^2 2 \theta_- \, \sin^2 \frac{\Delta m^{2}_{-}}{4k} t  \right. 
\nonumber \\ 
& + & \left. 2 \sin 2 \theta_+ \, \sin 2 \theta_- \,
\sin \frac{\Delta m^{2}_{+}}{4k} t \; 
\sin \frac{\Delta m^{2}_{-}}{4k} t \,
\cos \frac{\Sigma }{4k} t 
\right)
\label{24}
\eea
\bea
P(\nel \rightarrow  \nu_{\mu L}^c ) & = &  \frac{1}{4}  \left( 
 \sin^2 2 \theta_+ \, \sin^2 \frac{\Delta m^{2}_{+}}{4k} t \; + \;
 \sin^2 2 \theta_- \, \sin^2 \frac{\Delta m^{2}_{-}}{4k} t  
\right. \nonumber \\
& - & \left. 2 \sin 2 \theta_+ \, \sin 2 \theta_- \,
\sin \frac{\Delta m^{2}_{+}}{4k} t \;
\sin \frac{\Delta m^{2}_{-}}{4k} t \,
\cos \frac{\Sigma }{4k} t 
\right)
\eea

\bea
P(\nel \rightarrow  \nu^c_{eL} ) & = &  c^2_+ c^2_- 
\sin ^2 \, \frac{\Sigma - \Delta m_+^2 + \Delta m_-^2}{8 k} t  \, + \,
c^2_+ s^2_- 
\sin ^2 \, \frac{\Sigma - \Delta m_+^2 - \Delta m_-^2}{8 k} t  \nonumber \\
& + & s^2_+ c^2_- 
\sin ^2 \, \frac{\Sigma + \Delta m_+^2 + \Delta m_-^2}{8 k} t  \, + \,
s^2_+ s^2_- 
\sin ^2 \, \frac{\Sigma + \Delta m_+^2 - \Delta m_-^2}{8 k} t  \nonumber \\
& - & c^2_+ s^2_+ 
\sin ^2 \, \frac{\Delta m_+^2}{4 k} t  \, - \, c^2_- s^2_- 
\sin ^2 \, \frac{\Delta m_-^2}{4 k} t
\eea
and for the survival probability
\bea
P(\nel \rightarrow \nel) \; = \; 1 & - & c^2_+ c^2_- 
\sin ^2 \, \frac{\Sigma - \Delta m_+^2 + \Delta m_-^2}{8 k} t  \, - \,
c^2_+ s^2_- 
\sin ^2 \, \frac{\Sigma - \Delta m_+^2 - \Delta m_-^2}{8 k} t  \nonumber \\
& - & s^2_+ c^2_- 
\sin ^2 \, \frac{\Sigma + \Delta m_+^2 + \Delta m_-^2}{8 k} t  \, - \,
s^2_+ s^2_- 
\sin ^2 \, \frac{\Sigma + \Delta m_+^2 - \Delta m_-^2}{8 k} t  \nonumber \\
& - & c^2_+ s^2_+ 
\sin ^2 \, \frac{\Delta m_+^2}{4 k} t  \, - \, c^2_- s^2_- 
\sin ^2 \, \frac{\Delta m_-^2}{4 k} t
\eea
(we have adopted the shorthand notation
$c_{\pm} \, = \, \cos \, \theta_\pm$, $s_{\pm} \, 
= \, \sin \, \theta_\pm$). These probabilities in general 
depend on 5 parameters of the underlying 
theory, that is 2 mixing angles $\theta_{\pm}$, and 3 mass parameters 
$ \Delta m^2_{\pm} \; = \; m^2_{2 \pm} \, - \, m^1_{2 \pm} $,
$ \Sigma \, = \, m^2_{1+} + m^2_{2+} - m^2_{1-} - m^2_{2-}  $.\\
Note that in the limit of zero mixing angles all transition
probabilities vanish except $P(\nel \rightarrow \nu_{e L}^c)$ for which 
we recover the Pontecorvo oscillation formula \cite{Pontecorvo}, 
\cite{TE}: in this limit only neutrino-antineutrino oscillations 
with no flavour change are possible with both Dirac and Majorana mass
terms.\\
The obtained results for the transition probabilities in the general 
case, eqs. (\ref{216})-(\ref{218}), are rather complicate and 
then difficult to 
analyze. In the following we discuss some very interesting particular 
cases obtained for peculiar forms of Dirac and Majorana mass matrices.\\

\subsection{Pure Dirac and pure Majorana neutrinos.}

For the two limiting cases
\be
 M_D \; \neq \; 0 \;\;\;\;\;\;\;\;\;\; M_M \; = \; 0 \; 
\label{220}
\ee
(pure Dirac neutrinos) and
\be
 M_D \; = \; 0 \;\;\;\;\;\;\;\;\;\; M_M \; \neq \; 0 \; 
\label{221}
\ee
(pure Majorana neutrinos) we have
\be
 \theta_{+} \; = \; \theta_{-} \;\;\;\;\;\;\;\;\;\; 
  M_{+}^2 \; = \; M_{-}^2 \; 
\label{222}
\ee
so that
\bea
P(\nel \rightarrow \nml) & = &
\sin^2 2 \theta \, \sin^2 \frac{ \Delta m^2 }{4k} t
\label{223} \\
P(\nel \rightarrow  \nu^c_{\mu L} ) & = & 0 
\label{224} \\
P(\nel \rightarrow  \nu^c_{e L} ) & = & 0 
\label{225}
\eea
As we know, in these frameworks we have no neutrino-antineutrino 
oscillations but only flavour transitions, for which we recover 
the standard results \cite{Vac}. Note that for pure Dirac 
and pure  Majorana neutrinos the result is the same, but this
holds only in the ultrarelativistic limit \cite{Cimento}.\\

\subsection{The cases of Dirac mixing and Majorana masses and vice-versa.}

In general, all the elements of the $M_D$ and $M_M$ mass matrices 
in ({\ref{21})} are non zero; however, we can constrain these by making some 
physical ansatz.\\
Looking at the quark sector of the Standard Model we note
that quarks are Dirac particles (obviously!) and flavour
(weak interacting) eigenstates are mixed to give the mass eigenstates.
In analogy, we can assume that also for neutrinos the Dirac mass 
matrix is responsible
for the mixing of the flavour eigenstates. Nevertheless, we know
that neutrinos, if massive, are much more light than the corresponding
quarks (see the limits reported in \cite{masses}) so that one can think 
that their masses 
(but not necessarily mixings) are generated by a Majorana mass term.
The most simple forms for $M_D$ and $M_M$ translating these two ansatz 
are then 
\be
M_D \; = \;  \left( \ba{cc} 0  & m^D_{e \mu} \\
                             m^D_{e\mu}  & 0   \ea  \right)
\;\;\;\;\;\;\;
M_M \; = \;  \left( \ba{cc} m^M_{ee}  & 0 \\
                             0 & m^M_{\mu\mu}    \ea  \right)
\label{226}
\ee
We now explore the implications of (\ref{226}). The first one is that
\be
 U_{-} \; = \; U_{+}^{\dag} 
\label{227}
\ee
\be
 M_{+}^{2} \; = \; M_{-}^{2} 
\label{228}
\ee
i.e. + states  and $-$ states have equal mass eigenvalues but + states
are mixed between them in a way just opposite to that of $-$ states.
As a consequence\\
\be
 \theta_{-} \; = \; - \, \theta_{+} 
\label{229}
\ee
\be
 \Delta m_{+}^{2} \; = \; \Delta m_{-}^{2} 
\label{230}
\ee        
\be
 \Sigma \; = \; 0
\label{231}
\ee
Inserting these in the expressions for the transition probabilities 
(\ref{216})-(\ref{218}) we get
\bea
P(\nel \rightarrow \nml) & = & 0 
\label{232} \\
P(\nel \rightarrow  \nu^c_{\mu L} ) & = & 
\sin^2 2 \theta_+ \, \sin^2 \frac{ \Delta m^2_+ }{4k} t
\label{233} \\
P(\nel \rightarrow  \nu^c_{e L} ) & = & 0 
\label{234}
\eea
We see that, in this case, pure flavour oscillations and pure 
neutrino-antineutrino (Pontecorvo) oscillations are not allowed,
while only flavour changing neutrino-antineutrino transitions are predicted.
It is interesting to note that for the latter, the expression
for the transition probability has the same form as (\ref{223}).
As we will remark below, this has implications on the interpretation
of disappearance neutrino oscillation experiments.

On the contrary to the ansatz just analyzed, we can further explore 
the possibility that neutrino mixing is given only by the Majorana
mass term, while masses are generated directly by the Dirac term.
We then consider the following mass matrices:
\be
M_D \; = \;  \left( \ba{cc} m^D_{ee}  & 0 \\
                             0   & m^D_{\mu\mu}    \ea  \right)
\;\;\;\;\;\;\;
M_M \; = \;  \left( \ba{cc} 0  & m^M_{e \mu} \\
                             m^M_{e\mu}  & 0    \ea  \right)
\label{235}
\ee
Also in this case we find that relations (\ref{227}),(\ref{228}) hold, 
so that again we have eqs. (\ref{232})-(\ref{234}) 
for the transition probabilities.
The fact that both cases analyzed in this paragraph lead to the
same phenomenological predictions is analogous to that encountered 
in the previous paragraph, where (\ref{220}) and (\ref{221}) 
also gave identical 
results. We then observe a symmetry between Dirac and Majorana mass 
terms.\\

\subsection{ The case of degenerate Dirac and  Majorana mixing}

Another interesting ansatz is to suppose that neutrino mixing is 
generated by Dirac and Majorana mass terms with the same strength.
This can be simply implemented by using
\be
M_D \; = \;  \left( \ba{cc} m^D_{ee}  & m_{e \mu} \\
                             m_{e\mu}  & m^D_{\mu\mu}    \ea  \right)
\;\;\;\;\;\;\;
M_M \; = \;  \left( \ba{cc} 0  & m_{e \mu} \\
                             m_{e\mu}  & 0    \ea  \right)
\label{236}
\ee
or
\be
M_D \; = \;  \left( \ba{cc} 0  & m_{e \mu} \\
                             m_{e\mu}  & 0    \ea  \right)
\;\;\;\;\;\;\;
M_M \; = \;  \left( \ba{cc} m^M_{ee}  & m_{e \mu} \\
                             m_{e\mu}  & m^M_{\mu\mu}    \ea  \right)
\label{237}
\ee
In both cases we have one non-vanishing mixing angle and 2
(nearly) independent mass parameter; more precisely
\be
 \theta_{-} \; = \; 0
\label{238}
\ee
\be
 \Delta m_{-}^{2} \; = \; \cos 2 \theta_{+} \Delta m_{+}^{2} 
\label{239}
\ee
while $\Sigma$ is given by
\be
\Sigma \; = \; \frac{2 \, \sin^2 \, 2 \theta_+ \: \left( \Delta m_+^2 
\right)^2}{\left( m_{1-} \; + \; m_{2-} \right)^2}
\label{239b}
\ee
Note that in the limit of zero mixing angle we are left with only one 
mass parameter ($\Sigma = 0$). Instead for arbitrary mixing, in the 
present case, if the states $\nu_{1+}$, $\nu_{2+}$ are degenerate in 
mass ($\Delta m_+^2 \, = \, 0$) then {\it all} the four states
$\nu_{1 \pm}$, $\nu_{2 \pm}$ are degenerate ($\Delta m_+^2 \, = 
\, \Delta m_-^2 \, = \, \Sigma \, = \, 0$) so that no transition can 
occur. \\
Inserting (\ref{238}), (\ref{239})
in (\ref{216})-(\ref{218}) we now observe that all the 
transition probabilities are different from zero, and are given by
\be
P(\nel \rightarrow \nml) \; = \;
P(\nel \rightarrow  \nu^c_{\mu L} ) \; = \;
\frac{1}{4} \sin^2 2 \theta_+ \sin^2 \frac{ \Delta m^2_+ }{4k} t ;
\label{240} 
\ee
\bea
P(\nel \rightarrow  \nu^c_{e L} ) & = &
c^2_+ \sin^2 \left( \frac{\Sigma - 2 s_+ \Delta m^2_+}{8k} t \right) \; 
+ \; s^2_+ \sin^2 \left( \frac{\Sigma + 2 c_+ \Delta m^2_+}{8k} t \right) 
\\
& - & c^2_+ s^2_+ \sin^2 \left( \frac{\Delta m^2_+}{4k} t \right)
\label{242}
\eea
Interestingly, let us note that both $ \nel \rightarrow \nml $ and 
$\nel \rightarrow \nu^c_{\mu L} $ transitions 
have the same probability, whose form is identical to (\ref{223}) and 
(\ref{233}) apart a constant suppression factor for the oscillation 
amplitude of 1/4. For degenerate Dirac-Majorana mixing we then predict 
non vanishing flavour-conserving oscillations and equal 
probabilities for flavour-changing ones.

\section{Dirac-Majorana neutrino oscillations in matter}

In this section we generalize the MSW theory \cite{MSW} to take into 
account the Dirac-Majorana nature of neutrinos. \\
Let us consider neutrinos travelling in a medium with constant 
density, whose electron and neutron number densities are given by 
$N_e$ and $N_n$, respectively. In the flavour basis, the evolution 
equations are simply given by (\ref{23}) where the energy $\omega$ is 
substituted by $\omega \, - \, V$, $V$ being the effective potential 
experienced by a given neutrino state in the medium, namely 
\cite{MSW, Capone, Cimento}
\bea
V_{\nu_{eL}} & = & - \; V_{\nu^c_{eR}} \; = \; \sqrt{2} \, G_F \, 
\left( N_e \, - \, \frac{1}{2} \, N_n \right)  \label{31} \\
V_{\nu_{\mu L}} & = & - \; V_{\nu^c_{\mu R}} \; = \; - \, 
\frac{G_F}{\sqrt{2}} \, N_n \label{32} \\
V & = & 0 \;\;\;\;\;\;\;\;\;\;\;\;\;\; \mathrm{for \; all \; the \; other 
\; sterile \; states} \label{33}
\eea
Here $G_F$ is the Fermi coupling constant, and for simplicity we 
consider only non magnetized media (the generalization to these media 
is straightforward following \cite{Capone}). In compact form, the 
evolution equations are then given by
\be
\left( \ba{cc} k \, + \, V_L & M \\ M & - \, k \, - \, V_R 
\ea \right) \left( \ba{c} n_L \\ 
n_R \ea \right) \; = \; \omega \, \left( \ba{c} n_L \\ 
n_R \ea \right)
\label{34}
\ee
where
\bea
V_L & = &  diag \left\{ V_{\nu_{eL}}, \, V_{\nu_{\mu L}}, \, 0 , \, 0
\right\} \label{35} \\
V_R & = & diag \left\{ 0, \, 0, \, - \, V_{\nu^c_{eR}}, \, - \, 
V_{\nu^c_{\mu R}} \right\} \label{36}
\eea
In the ultrarelativistic limit, and for $V_{\nu_{eL}}, \, V_{\nu_{\mu 
L}} \ll k$, we can again separate the evolution of $n_L$ and $n_R$ 
states, and in particular we get
\be
H_m \, n_L \; = \; \omega \, n_L
\label{37}
\ee
with the effective hamiltonian given by
\be
H_m \; \simeq \; k \; + \; \frac{M^2}{2 \, k} \; + \; V_L
\label{38}
\ee
In terms of the physical mass and mixing parameters, subtracting from 
$H_m$ terms proportional to the identity matrix which contribute with 
an irrelevant common phase factor to the wave functions, the effective 
hamiltonian in the flavour basis can be explicitly written as
\be
H_m \; = \; \frac{1}{4} \, \left( \ba {cccc}
- \, D_{+c} \, + \, 4 \, V_{\nu_{eL}} 
& - \, D_{+s}  & \sigma \, - \, D_{-c} & - \, D_{-s} \\
- \, D_{+s} & D_{+c} \, + \, 4 \, V_{\nu_{\mu L}}  & 
- \, D_{-s} & \sigma \, + \, D_{-c} \\
\sigma \, - \, D_{-c} & - \, D_{-s} & 
- \, D_{+c} & - \, D_{+s} \\
- \, D_{-s} & \sigma \, + \, D_{-c} &
- \, D_{+s} & D_{+c}
\ea \right) \label{39}
\ee
where we have used the notations 
\bea
D_{\pm c} & \simeq &  \frac{\Delta m_+^2}{2 k} \, c_{2+} \, \pm \, 
\frac{\Delta m_-^2}{2 k} \, c_{2-} \\ 
D_{\pm s} & \simeq &  \frac{\Delta m_+^2}{2 k} \, s_{2+} \, \pm \, 
\frac{\Delta m_-^2}{2 k} \, s_{2-} \\
\sigma & = & \frac{\Sigma}{2 k}
\eea
Diagonalizing $H_m$ in (\ref{39}) we then get the matter mass 
eigenstates from which the transition probabilities in matter can be 
obtained. Before dealing with this point, we first want to discuss the 
occurrence of resonances in neutrino matter oscillations.

\subsection{Resonance conditions}

Resonant enhancement of oscillations takes place when two unperturbed 
(mixing equals zero) energy levels cross between them \cite{MSW}. The 
resonance conditions can then be (approximately) obtained by equating 
the diagonal elements of $H_m$ in (\ref{39}).\\
For resonant \nel \rt \nml transitions we have the following condition
\be
\frac{\Delta m_+^2}{2 \, k} \, \cos \, 2 \theta_+ \; + \; 
\frac{\Delta m_-^2}{2 \, k} \, \cos \, 2 \theta_- \; = \; 2 \, 
\sqrt{2} \, G_F \, N_e
\label{310}
\ee
while for \nel \rt \ncml  and \nml \rt \ncel
\bea
\frac{\Delta m_+^2}{2 \, k} \, \cos \, 2 \theta_+ \; + \; 
\frac{\Delta m_-^2}{2 \, k} \, \cos \, 2 \theta_- & = & 2 \, 
\sqrt{2} \, G_F \, \left( N_e \, - \, \frac{1}{2} \, N_n \right)
\label{311} \\
\frac{\Delta m_+^2}{2 \, k} \, \cos \, 2 \theta_+ \; + \; 
\frac{\Delta m_-^2}{2 \, k} \, \cos \, 2 \theta_- & = & 
\sqrt{2} \, G_F \, N_n
\label{312}
\eea
respectively. Instead, as already found in \cite{previous}, the 
active-sterile flavour-conserving matter transitions \nel \rt \ncel 
and \nml \rt \ncml have a maximum amplitude only if 
\be
V_{\ne} \, = \; 0 \label{313}
\ee
or 
\be
V_{\nm} \; = \; 0 \label{314}
\ee
respectively \footnote{Note that, while eq. (\ref{314}) is realized 
only in vacuum, relation (\ref{313}) can be satisfied also in a medium 
with $N_e \, = \, N_n / 2$. Such a condition can be achieved in the 
first stages of the neutronization phase of a neutron star.}. \\
The relation (\ref{310}) generalizes to Dirac-Majorana neutrinos the 
resonance condition for flavour oscillations obtained in \cite{MSW} 
for pure Dirac or pure Majorana (ultrarelativistic) neutrinos:
\be
\frac{\Delta m^2}{2 \, k} \, \cos \, 2 \theta \; = \; \sqrt{2} \, G_F 
\, N_e
\label{315}
\ee
An important novel feature emerging from (\ref{310}) is that, on the 
contrary to what happens for pure Dirac or pure Majorana neutrinos 
(eq. (\ref{315})), the resonance condition is ruled by {\it two} 
squared masses differences so that, even if two or three of four mass 
eigenvalues are degenerate, the enhancement of oscillations can still 
take place (note that the resonance density in (\ref{310}) is shifted 
with respect to that occurring for (\ref{315}) towards lower values). 
The same is valid for active-sterile (flavour-changing) transitions.
Note, however, that the resonance conditions do not depend on the 
$\Sigma$ parameter.

For the particular case in which the Dirac and Majorana mass matrices 
are given by (\ref{226}) or (\ref{235}), only one transition (\nel \rt 
\ncml) can occur, and for this the resonance condition reduces to
\be
\frac{\Delta m^2_+}{2 \, k} \, \cos \, 2 \theta_+ \; = \; \sqrt{2} \, G_F 
\, \left( N_e \, - \, \frac{1}{2} \, N_n \right)
\label{315b}
\ee
Instead, for the mass matrices taking the form in (\ref{236}) or 
(\ref{237}) all the transition discussed in the general case can be 
resonant, but now they are all ruled by only one squared masses 
difference. So, the resonance condition for \nel \rt \nml
\be
\frac{\Delta m^2_+}{2 \, k} \, \cos \, 2 \theta_+ \; = \; \sqrt{2} \, G_F 
\, N_e
\label{316}
\ee
is phenomenologically equivalent to that for pure Dirac or pure 
Majorana neutrinos, while the one for \nel \rt \ncml is again given by 
(\ref{315b}).

After this qualitative discussion on the resonant enhancement of 
matter oscillations, we now proceed to find the expressions for the 
transition probabilities. As for the vacuum, this can be achieved 
also in the general case (described by the hamiltonian $H_m$ in 
(\ref{39})) because the eigenvalue equation corresponds to a fourth 
degree algebraic equation whose solutions are known analytically. 
However, the resulting expressions for the probabilities  are very 
involved and not very informative as (and much more than) in the 
vacuum case. So we will only consider the particular scenarios 
envisaged in the previous section and for these we will single out the 
explicit form of the transition probabilities.

\subsection{Pure Dirac and pure Majorana neutrinos}

Given (\ref{220}) or (\ref{221}) the problem is separable into the 
diagonalization of two $2 \times 2$ effective hamiltonians, one for 
the active left-handed neutrino states and the other for the sterile 
left-handed antineutrino states (which propagate freely as in vacuum) 
for the Dirac case or for the active right-handed antineutrino states 
for the Majorana case. In both cases we have
\be
H_m \, \nu_L \; = \; \omega \, \nu_L
\label{317}
\ee
with
\be
H_m \; \simeq \; k \; + \; \frac{M^2_{D,M}}{2 \, k} \; + \; \left( 
\ba{cc} V_{\nel} & 0 \\ 0 & V_{\nml} \ea \right)
\label{318}
\ee
and so we recover the standard MSW theory \cite{MSW}. \\
No active-sterile (flavour-changing) transition occurs in this case, 
as in vacuum.

\subsection{The cases of Dirac mixing and Majorana masses and 
vice-versa}

Let us now assume that mass matrices are given by (\ref{220}) or 
(\ref{235}) and substitute relations (\ref{229})-(\ref{231}) in the 
effective hamiltonian (\ref{39}). Inspired from the results obtained 
for the vacuum case, it is useful to introduce the permutated flavour 
basis
\be
\left( \ba{c} \nel \\ \ncml \\ \ncel \\ \nml \ea \right)
\label{319}
\ee
In this basis, the problem is again separable because the effective 
hamiltonian becomes block-diagonal. We then have
\bea
H_m^{(1)} \, \left( \ba{c} \nel \\ \ncml \ea \right) & = & \omega \, 
\left( \ba{c} \nel \\ \ncml \ea \right) \label{320} \\
H_m^{(2)} \, \left( \ba{c} \ncel \\ \nml \ea \right) & = & \omega \, 
\left( \ba{c} \ncel \\ \nml \ea \right) \label{321}
\eea
with
\bea
H_m^{(1)} & \simeq & k \; + \; \frac{\cal{M}^2}{2 \, k} \; + \; \left( 
\ba{cc} V_{\nel} & 0 \\ 0 & 0 \ea \right) \label{322} \\
H_m^{(2)} & \simeq & k \; + \; \frac{\cal{M}^2}{2 \, k} \; + \; \left( 
\ba{cc} 0 & 0 \\ 0 & V_{\nml} \ea \right) \label{323} 
\eea
where the mass matrix $\cal{M}$ is
\be
\cal{M} \; = \; \left( \ba{cc} m^D_{ee} & m^M_{e \mu} \\ m^M_{e \mu} & 
m^D_{\mu \mu} \ea \right)
\label{324}
\ee
Let us focus, for example, on (\ref{322}). It has the same form of the 
$2 \times 2$ 
effective $e - \mu$ MSW hamiltonian (\ref{318});
the expressions for the transition probabilities in matter are then
\bea
P(\nel \rightarrow  \nml) & = & 0 
\label{325} \\
P(\nel \rightarrow \ncml) & = &
\sin^2 2 \theta_+^m \,  \sin^2 \frac{ \pi x}{L_m} 
\label{326} \\
P(\nel \rightarrow  \ncel) & = & 0 
\label{327}
\eea
($x\simeq t$) where the effective mixing angle in matter is given by
\be
\sin \, 2 \theta^m_+ \; = \; \frac{ \frac{\Delta m_+^2}{2k} \, \sin \, 2 
\theta_+}{\sqrt{\left( \frac{\Delta m^2_+}{2k} \, \cos \, 2 
\theta_+ \; - \; \sqrt{2} \, G_F \, (N_e \, - \, \frac{1}{2} \, N_n ) 
\right)^2 \; + \; \left( \frac{\Delta m_+^2}{2k} \, \sin \, 2 
\theta_+ \right)^2}}
\label{328}
\ee
while the effective oscillation length by
\be
L_m \; = \; \frac{2 
\pi}{\sqrt{\left( \frac{\Delta m_+^2}{2k} \, \cos \, 2 
\theta_+ \; - \; \sqrt{2} \, G_F \, (N_e \, - \, \frac{1}{2} \, N_n ) 
\right)^2 \; + \; \left( \frac{\Delta m_+^2}{2k} \, \sin \, 2 
\theta_+ \right)^2}}
\label{329}
\ee
From (\ref{328}) we then obtain again that the transitions \nel \rt 
\ncml are resonantly amplified if the resonance condition (\ref{315b}) 
is fulfilled.\\
At this point we stress the fact that for Dirac-Majorana neutrinos 
described by the mass matrices in (\ref{220}) or (\ref{235}) the MSW 
theory \cite{MSW} applies practically unmodified to \nel \rt \ncml 
instead of \nel \rt \nml flavour transitions, and so the 
phenomenological implications for disappearance experiments are the 
same for the two cases. In particular, the analysis performed in 
\cite{Stsol} for the solar neutrino problem holds true for the present 
case, and from that the values of $\Delta m^2_+$ and $\sin^2 \, 2 
\theta_+$ able to solve the puzzle can be extracted.

\subsection{The case of degenerate Dirac-Majorana mixing}

Let us now turn on the last model considered in the previous section, 
with the mass matrices given by (\ref{236}) or (\ref{237}). From 
(\ref{238}), (\ref{239}) we then find that the proper effective 
hamiltonian to diagonalize is
\bea
H_m & = & \frac{1}{8 k} \, \left( \ba{cccc}
- \, 2 \, \Delta m_+^2 \, \cos \, 2 \theta_+ 
& - \, \Delta m_+^2 \, \sin \, 2 \theta_+ &
\Sigma & - \, \Delta m_+^2 \, \sin \, 2 \theta_+  
\\
- \, \Delta m_+^2 \, \sin \, 2 \theta_+ & 
2 \, \Delta m_+^2 \, \cos \, 2 \theta_+ 
& - \, \Delta m_+^2 \, \sin \, 2 \theta_+ & 
\Sigma \\
\Sigma & - \, \Delta m_+^2 \, \sin \, 2 \theta_+ 
& - \, 2 \, \Delta m_+^2 \, \cos \, 2 \theta_+ & 
- \, \Delta m_+^2 \, \sin \, 2 \theta_+ \\
- \, \Delta m_+^2 \, \sin \, 2 \theta_+ & \Sigma 
& - \, \Delta m_+^2 \, \sin \, 2 \theta_+ &
2 \, \Delta m_+^2 \, \cos \, 2 \theta_+ \ea \right) \; + \nonumber \\
& \left. \right. & \left. \right. \nonumber \\
& + & diag\{ V_{\nu_{eL}} , V_{\nu_{\mu L}} , 0 , 0 \}
\label{q1}
\eea
Because of the presence of $V_{\nu_{\mu L}}$, we observe that the 
submatrices corresponding to the subsystem $(\nu_{eL} , \nu_{\mu L})$ 
and $(\nu_{eL} , \nu_{\mu L}^c)$ are now not equal (as instead 
happened for the vacuum case) so that \nel \rt \nml and \nel \rt \ncml 
oscillations have different transition probabilities. The interactions 
with the medium remove this sort of degeneracy; this was manifest 
already in the expressions for the resonance conditions (\ref{310}) and 
(\ref{311}).\\
The exact eigenvalue problem for $H_m$ in (\ref{q1}) involves again 
complicated solutions of a fourth-degree equation. Instead of giving 
the general expressions for the transition probabilities, it is more 
useful to discuss qualitatively the oscillations pattern which is very 
similar to the general one described by $H_m$ in (\ref{39}) since in 
both cases all the transition between the different flavour states can 
take place (the main differences between the two cases will be 
remarked in the following section). \\
For a medium with varying density we can qualitatively analyze 
neutrino propagation in it with the help of a level crossing diagram, 
depicted (for the present case) in Fig. 1. Full lines represent the 
eigenvalues of $H_m$ in (\ref{q1}) plotted against the medium density, 
while the dashed lines correspond to the unperturbed (no mixing) 
energy eigenvalues (i.e. to the diagonal elements of $H_m$ with 
$\theta_+ \, = \, 0$, since the non diagonal elements all vanish in 
the zero mixing limit). The crossing points of the unperturbed levels 
(approximatively) identify the resonance densities $\rho_{R i}$
 for the transitions 
\nel \rt \nml (R1), \nml \rt \ncel (R2) and \nel \rt \ncml (R3). The 
relative positions of the three resonance points depend on the $Y_e \, 
= \, Z/A$ ratio of the considered medium; as one can see from 
(\ref{310})-(\ref{312}),
\bea
\rho_{R1} & \sim & Y_e^{-1}  \label{q2} \\
\rho_{R2} & \sim & \left| \frac{1 \; - \; Y_e}{2} \right|^{-1} 
\label{q3} \\
\rho_{R3} & \sim & \left| \frac{3 Y_e \; - \; 1}{2} \right|^{-1} 
\label{q4}
\eea
In Fig. 1 we have chosen $Y_e \, = \, 0.48$; in the particularly 
common case in which $Y_e \, = \, 0.5$ we would have that both R2 and 
R3 occur at the same density. \\
Let us now focus on the evolution od Dirac-Majorana neutrinos 
propagating in a varying density medium, such as for example the Sun 
or another star like this. Neutrinos are produced deep in the star, at 
high densities, in flavour eigenstates (typically \nel) and then move 
out towards low density regions. As can be seen from Fig. 1, in fact, at high 
densities mixing effects can be neglected and the energy eigenvalues 
practically coincide with the unperturbed ones. The subsequent 
evolution, with decreasing density, depends whether the resonances are 
crossed  adiabatically or not (i.e., qualitatively, if neutrinos 
travelling in the medium ``see'' a density which varies very slowly 
during their path or not). If the adiabaticity condition \cite{Kuo} is 
fulfilled (at each resonance), neutrinos evolve according to the 
unbroken lines; otherwise at a given resonance there is a non 
vanishing probability for ``jumping'' from one level to another one 
\cite{Kuo}. So, following Fig. 1, if for example we have a \nel 
$\simeq \nu_4$ at high densities and the resonance R3 is crossed 
adiabatically, at intermediate densities we again encounter a $\nu_4$ 
but now $\nu_4 \simeq \ncml$: we have had a transformation of an 
active \nel into a sterile \ncml. Instead, if R3 is crossed non 
adiabatically so that there is a non zero probability to jump onto the 
$\nu_1$ state, at intermediate densities we encounter (with a certain 
probability) a $\nu_1 \simeq \nel$ and practically we have had no 
conversion. But moving towards lower density another resonance point 
(R1) is present: if this is crossed adiabatically, then after that we 
have a $\nu_1 \simeq \nml$ (active-active \nel \rt \nml conversion), 
otherwise we have a jump onto the $\nu_2$ state which (after the 
resonance) is approximatively a \nel (no conversion). \\
Confronting the present scenario with that of the usual MSW theory, we 
see that now efficient active-active flavour-changing conversions can 
be obtained only if the resonance for active-sterile transitions (R3 
or R2) is crossed non adiabatically while that for active-active 
transition (R1) is crossed adiabatically. But the important novel 
feature regarding Dirac-Majorana neutrinos is that exiting from the 
medium we do {\it not} have pure flavour states. In fact, as one can 
see from Fig. 1 and contrarily to what happens for the standard MSW 
theory, at very low density (approaching the vacuum) the energy 
eigenstates (solid lines) do not approach flavour eigenstates (dashed 
lines). This is a genuine feature of the Dirac-Majorana nature of 
neutrinos and it can be shown that (in the present case)
this is due to the fact that in the 
limit of zero mixing the $\Sigma$ parameter vanishes (see eq. 
(\ref{239b})) \footnote{In fact, keeping $\Sigma$ fixed in the limit 
$\theta_+ \rt 0$, at very low density the dashed lines approach the 
solid ones}. \\
However, this is not a completely surprising feature, because we 
already know that in vacuum even for zero mixing the pure flavour 
eigenstates are not the physical eigenstates, which are instead given 
by the Majorana combinations $\wt{n_{\pm}}$ in (\ref{212}). The level 
crossing diagram in Fig. 1 for zero density is just an expression of 
this physical fact. For zero mixing we have only two (doubly 
degenerate) energy eigenvalues $\pm \frac{\Delta m_+^2}{4k}$ while, 
switching on the mixing, four non degenerate energy eigenvalues 
appear,
\be
\frac{1}{8k} \, \left( \pm \, 2 \Delta m_+^2 \; + \; \Sigma \right)
\label{q5}
\ee
\be
\frac{1}{8k} \, \left( \pm \, 2 \Delta m_+^2 \, \cos \, 2 \theta_+ 
\; - \; \Sigma \right) \label{q6}
\ee
and the physical eigenstates at the exit of the medium are just 
$\nu_1$, $\nu_2$, $\nu_3$, $\nu_4$ $=$ $\wt{n_{\pm}}$ corresponding to 
the energy values given in (\ref{q5}), (\ref{q6}).

\section{Concluding remarks}

In this paper we have studied the propagation both in vacuum and 
in matter of Dirac-Majorana neutrinos and analyzed 
active-active (flavour-changing) as well as active-sterile
transitions, which are, in general, both possible.\\
For vacuum oscillations, in section 2 we have given the general 
expressions for the transitions probabilities for $ \nel \rightarrow \nml $,
\nel \rt \ncml , $ \nel \rightarrow \ncel $
We have then discussed some interesting limiting cases for Dirac
($M_D$) and Majorana ($M_M$) mass matrices. For pure Dirac 
($M_M=0$) or 
pure Majorana ($M_D = 0$) neutrinos obviously we recover the usual
flavour oscillation formulae \cite{Vac}, while for both
$M_D$ and $M_M$ non vanishing and diagonal the Pontecorvo formula
\cite{Pontecorvo} for neutrino-antineutrino (active-sterile) 
oscillations is obtained \cite{TE}.\\
An interesting non trivial case is that with $M_D$ and $M_M$ given by
(\ref{226}) or (\ref{235}) which implement the idea that neutrino mixing is 
essentially ruled only by the Dirac mass term while the Majorana 
mass term is diagonal or vice-versa, respectively.
In both cases, neither pure flavour oscillations nor Pontecorvo
oscillations are predicted, but only flavour-changing active-sterile 
transitions, such as $\nel \rightarrow \ncml$, 
are possible. Remarkably, the transition probability
for these is identical in form to that for flavour oscillations 
for pure Dirac or Majorana neutrinos, and this holds both in vacuum and 
in matter. For the latter, the resonance condition is only 
shifted by the neutral current contribution of \nel to the 
effective potential. So, for example, the solution to the solar 
neutrino problem in terms of active-sterile neutrino oscillations 
proposed in \cite{Stsol} applies unmodified to the present scheme. 
\\
Another interesting case, even if a bit more complicate, has been
analysed for the mass matrices in (\ref{236}) or (\ref{237}), which implements
the idea that neutrino mixing is given by the Dirac and Majorana
mass terms with the same strength. In this case, $ \nel \rightarrow \nml $,
$ \nel \rightarrow \ncml $, $ \nel \rightarrow \ncel $ transitions
are all possible and, in vacuum, the first two have the same
transition probability, which is also identical in form to that
obtained in the previous case, except for a constant suppression 
factor in the amplitude of oscillations of 1/4. Also in matter
the pattern of neutrino transitions present in the general case is 
(qualitatively) reproduced in this peculiar scheme. In particular,
all the flavour changing oscillations can be 
resonantly amplified while Pontecorvo \nel \rt \ncel matter 
oscillations have maximum amplitude only for a given 
electron to neutron number ratio (see eq. (\ref{313}) and the related 
footnote); the resonance conditions were discussed in section 3.1 . \\
Given the multiresonance structure of the oscillations pattern, it is 
then interesting to follow the evolution of a \nel , for example, in a 
varying density medium such as the Sun; this has been done in section 
3.4 with the help of a level crossing diagram reported in Fig. 1. 
Several scenarios are possible according to the adiabaticity 
properties of level crossing near the resonance points. In particular, 
starting from a pure \nel beam at high density, to have a consistent 
conversion into \nml at low density the resonance for \nel \rt \ncml 
has to be crossed non adiabatically, while the passage through the one 
for \nel \rt \nml has to be adiabatic. However, we have also shown 
that at very low density, and then in the vacuum, it is more 
appropriate to deal with the Majorana combinations $\wt{n_\pm}$ in 
(\ref{212}) than with the pure flavour states \nel , \ncel , \nml , 
\ncml . This is strictly related to the Dirac-Majorana nature of 
neutrinos, which chooses Majorana eigenstates instead of pure flavour 
states as starting points. In this respect, we have to deal with 
``generic'' flavour-changing or flavour-conserving transitions of 
Dirac-Majorana neutrinos without looking at the particular active 
neutrino or sterile antineutrino state. It is through the weak 
interactions, with which neutrinos are produced and detected, that a 
particular (active or sterile) component of the Majorana eigenstates is 
chosen.\\
The results obtained for the case of degenerate Dirac-Majorana mixing 
are qualitatively valid also in the general case in which all the 
entries of the Dirac and Majorana mass matrices are non zero and 
different between them. The main difference between the two cases is 
that in the general framework there are 3 mass parameters and two 
mixing angles ruling the evolution, while for the particular case 
studied in sections 2.3 and 3.4 there are only 2 mass parameters and 1 
mixing angle (these parameters being not completely independent, 
because of relation (\ref{239b})). The presence of more degrees of 
freedom in the general case allows to consider some peculiar 
situations which are not possible otherwise. The most remarkable one 
is that in the general case the proportionality of the $\Sigma$ 
parameter to $\sin^2 \, 2 \theta_+$ (see eq. (\ref{239b})) is lost, so 
that the structure of the level crossing diagram at very low density 
can be altered. The eigenvalues of $H_m$ in (\ref{39}) for zero 
density (vacuum) are given by
\be
\frac{1}{8k} \, \left( \pm \, 2 \Delta m_+^2 \; + \; \Sigma \right)
\ee
\be
\frac{1}{8k} \, \left( \pm \, 2 \Delta m_-^2 \; - \; \Sigma \right) 
\ee
so that one can manipulate the mass parameters to modify the low 
density region of the level crossing diagram without grossly altering 
the region where the resonance points are present. In any case, there 
can be present no substantial modifications of the conclusions reached 
above.

The oscillations of Dirac-Majorana neutrinos here studied with their 
peculiar features can be efficiently tested in astrophysics, in 
particular detecting solar or supernova neutrinos, and can have even 
profound implications in cosmology for the nucleosynthesis of light 
elements in the Universe.

\vspace{1truecm}

\noindent
{\Large \bf Acknowledgements}\\
\noindent
We express our sincere thanks to Prof. F. Buccella for very useful
talks and his unfailing encouragement, and to Prof. E. Kh. Akhmedov 
for enlightening discussions with one of us (S.E.).

\newpage

\noindent

Figure 1: Level crossing diagram for 1 MeV momentum Dirac-Majorana 
neutrinos described by the mass matrices in (\ref{236}) or 
(\ref{237}). The eigenvalues of $H_m$ in (\ref{q1}) are plotted versus 
the density of a medium with $Y_e \, = \, Z/A \, = \, 0.48$. Neutrinos 
parameters are fixed as follows: $\Delta m_+^2 \, = \, 10^{-6} \, 
eV^2$, $\Sigma \, = \, 2.5 \times 10^{-7} \, eV^2$, $\sin \, 2 
\theta_+ \, = \, 0.1$. Dashed lines refer to the zero mixing limit 
($\theta_+ \, = \, 0$) of the energy eigenvalues.

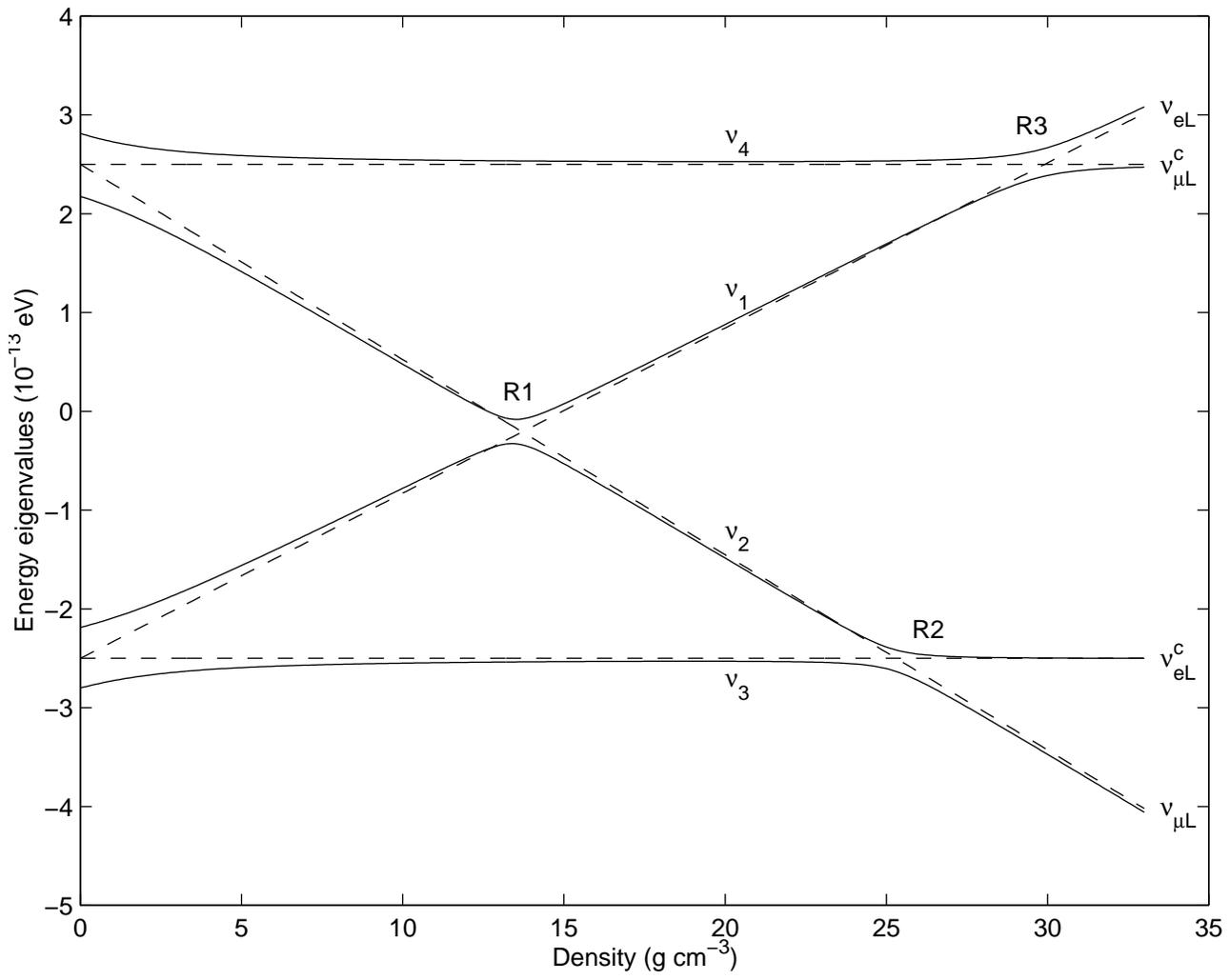
\end{document}